\documentclass[12pt,english]{article}
\usepackage[T1]{fontenc}
\usepackage[latin9]{inputenc}
\usepackage{color}
\usepackage{babel}

\usepackage{amssymb}
\usepackage[unicode=true, pdfusetitle,
 bookmarks=true,bookmarksnumbered=false,bookmarksopen=false,
 breaklinks=false,pdfborder={0 0 1},backref=false,colorlinks=true]
 {hyperref}

\makeatletter
\newcommand{\lyxaddress}[1]{
\par {\raggedright #1
\vspace{1.4em}
\noindent\par}
}

\makeatother

\begin{document}

\title{Free Fock space and functional calculus approach to the n-point information
about the {}``Universe''}

\author{Jerzy Han\'{c}kowiak}
\maketitle
\begin{abstract}
{\large Starting from a differential equation for the unique field
$\varphi(\tilde{x})$, where the vector $\tilde{x}$ contains space-time
and the discrete field characteristics, the equation for the generating
vector |V> of the n-point information (correlation and smeared functions)
in the free Fock space is derived. In derived equation, due to appropriate
extension of the right invertible operators, the physical vacuum vector
$|0>_{ph}$appears with a global characteristic of the field $\varphi$. }{\large \par}

{\large For so called resolvent regularization of the original systems,
the closed equations for the n-point information are analysed with
the help of functional calculus. }{\large \par}

{\large key words: the strong and a weak formulation, a right invertible
operator, a regularized or modified theory, a generalized resolvent,
functional calculus}{\large \par}
\end{abstract}

\lyxaddress{{\large Zielona Gora University, e-mail: hanckowiak@wp.pl}}

\section{{\large Introduction}}

{\large In spite of a growing} {\large trend in physics to define
the physical world as being made of information itself and thus information
is defined in this way, I am using this term to express some knowledge
about things and ideas. To gain this information - before you need
to formulate a suitable question, see Titus Lucretius Carus }\textit{\large De
Rerum Natura.}{\large{} According to Britannica Concise Encyclopaedia
- equations, in essence, - are questions. To get corresponding knowledge
(information about things and ideas) we have to solve these equations
which in many cases is not an easy task. The trouble is that the questions
are too detailed and hence the idea of }\textit{\large weak formulation}{\large{}
of the original equations is used. In Internet we can find the following
characterization of this idea:}{\large \par}

{\large {}``Weak formulations are an important tool for the analysis
of mathematical equations that permit the transfer of concepts of
linear algebra }\textbf{\large to solve problems in other fields}{\large{}
such as partial differential equations. In a weak formulation, an
equation is no longer required to hold absolutely (and this is not
even well defined) and has instead }\textit{\large weak solutions
}{\large only with respect to certain \textquotedbl{}test vectors\textquotedbl{}
or \textquotedbl{}test functions\textquotedbl{}''. See Wikipedia
<Weak formulation>; the page last modified on 27 January 2010 by unknown
author?. }{\large \par}

{\large An example in which the idea of weak formulation is used is
a celebrated Galerkin method. In this method the original equations
are not changed but the original spaces in which solutions are searched
are drastically changed, for example, when the original space is substituted
by a finite dimensional usually a low dimensional subspace. Surprising
is that in this way in many cases you can get quite correct results
even for very complicated systems describing, for example, the fluid
flows, see, e.g.,\cite{Noak (2010)}, \cite{Cordier (2009)}, \cite{Montlaur (2009)}
and Internet. }{\large \par}

{\large We have to remember, however, that the weak solutions of the
weak equations, obtained in a frame of reduced-order philosophy, are
not solutions of the original problem, see \cite{Gunzburger (????)}. }{\large \par}

{\large In the paper presented, in contrary to the canonical situation,
we propose such use of the weak formulation of the original theory
that the weak solutions have a clear physical interpretation and perhaps
nice mathematical properies. So, instead of the original mostly partial
differential equations (PDE) with the initial and boundary conditions
(IBC) defined in a sharp way, we consider PDE in which IBC are trated
as random or smooth quantities. As a consequence of that approach,
the }\textbf{\large original}{\large{} }\textbf{\large nonlinear equations}{\large{}
are substituted by }\textbf{\large the linear equations}{\large{} for
correlation functions or their generalizations in both cases called
the n-point functions (n-pfs) or as in the title - the n-point information.
This step can be treated as a weak formulation of the original nonlinear
system of equations because considered linear equations contains solutions
of the original theory as well as the new solutions. The second step
in the paper is quite opposite to what is done in any }\textbf{\large canonical
weak formulation}{\large : instead of narrowing the space in which
n-point information are considered, we move to a larger space - the
free Fock space (in which we do not postulate the permutation symmetry
of functions). However, in this enlarged (free) Fock space - like
in a smaller space of weak formulation - the transfer of concepts
of linear algebra is possible and even general solutions in many cases
can be constructed. }{\large \par}

{\large Given the extraordinary ease of constructing unilaterally
reverse operations to many operators which appear in the free Fock
space and by introducing an additional parameter (minor coupling constant)
can be derived new equations for n-pfs. In this study and others we
were trying to better understand derived equations. }{\large \par}

{\large In Secs 2 and 3 we define the strong and weak formulations
of the considered equations.}{\large \par}

{\large In Sec.4 the free (super, general) Fock space metodolodgy
is described and the basic equation for the correlation functions,
(\ref{eq:4.5}) is postulated. }{\large \par}

{\large Sec.5 is devoted to the canonical perturbation theory applied
to the Eq.\ref{eq:4.5} and a determination of the arbitrary terms
which appear in the free Fock space, see also Sec.8. }{\large \par}

{\large It seems that very interesting is the idea of regularization
of the true culprit of many problems of nonlinear theory:}{\large \par}

{\large \begin{equation}
\lambda_{1}\hat{N}\rightarrow\lambda_{1}(\hat{I}+\lambda_{2}\hat{M})_{R}^{-1}\hat{N}\label{eq:1.1}\end{equation}
whereby in the derived equations in addition to the usual sum of operators
related to the linear ($(\hat{L}+\hat{G})$) and nonlinear ($\hat{N}$)
parts of the theory, see Eq.\ref{eq:2.1} and Eq.\ref{eq:4.5}, the
product ($\lambda_{2}(\hat{L}+\hat{G})\hat{M})$ appears, see Secs
7 and 8. In a particular case of regularization, $\hat{M}=\hat{N}$,
- the closed equations for n-pfs - are obtained. In Sec.9 remarks
about functional calculus are given. An example of such equations
and the general solution for the 1-pf in the case of the $\varphi^{3}$-
model is discussed in Sec.10 where some general remarks are also included. }{\large \par}

{\large This paper is a continuation of work \cite{Hanckow (2010)}
with , as we hope, better use of the operator-valued functions and
with useful extension of considered operators. In comparison with
work \cite{Hanckow (2007)}, where evolutionary type of equations
were considered, in this paper - equations of {}``resovent type''
are used, for comparison, see \cite{Ivan (1974)}. }{\large \par}

\section{{\large Strong (exact) formulation }}

{\large We will assume that a theory is formulated by the following
integro-differential equations}{\large \par}

{\large \begin{equation}
L[\tilde{x};\varphi(\tilde{x})]+\lambda N[\tilde{x};\varphi]+G(\tilde{x})=0\label{eq:2.1}\end{equation}
 with a linear and nonlinear dependence on the unique field $\varphi$
(first and second terms) and a free term G. In the case of homogeneous
environment or materials, the operators L and N do not depend explicitly
on the vector variable $\tilde{x}\in B$ where a set $B$ describes
the domain of the unique field $\varphi$ . In order to improve the
description of equations the components of the vector $\tilde{x}$
contain also discrete indexes which usually appear as subindices of
the fields or functions. It turns out that riddance of the lower and
upper indices is a big improvement of description. In this way only
one }\textit{\large unique field}{\large{} $\varphi$ is considered.
The coupling constant $\lambda$, later denoted by $\lambda_{1}$and
called the major coupling constant, contains the memory of the nonlinearity
of the original, strong formulation (\ref{eq:2.1}). This is usually
an expansion parameter in the perturbation approach to the statistical
and quantum fields. }{\large \par}

{\large We rewrite the Eq.\ref{eq:2.1} as }{\large \par}

{\large \begin{equation}
(L_{0}\varphi)(\tilde{x})+L_{1}[\tilde{x};\varphi(\tilde{x})]+\lambda N_{0}[\tilde{x};\varphi(\tilde{x})]+\lambda N_{1}[\tilde{x};\varphi]+G(\tilde{x})=0\label{eq:2.2}\end{equation}
This is an equation for the unique field $\varphi(\tilde{x})$ in
which operators (functionals) with sub index {}``o'' denote a local
or self interaction of a cell or particle or field, but sub index
{}``1'' denotes an interaction among the constituents of the physical
system. An additional restriction of particular terms in Eq.\ref{eq:2.2}
comes from a co-variant character of proposed equations with respect
to symmetry transformations of the theory. A symmetry of equations
can be used to introduce another averages (or smoothing) than the
ensemble averages with a simple, geometrical interpretation and often
direct measured. It is interesting that these two kind of smoothing
procedure lead to identical equations for n-pfs, \cite{han2007},\cite{han2008}.
This is a happy coincidence because in this way we can compare theory
with experiment without resolving the ergodic problem, \cite{badino2005}.
The problem of calculation of the time averages is related to solving
equations upon n-pfs with appropriate additional conditions. }{\large \par}

\section{{\large Weak formulation can be linear, multitime and noncommuting.
Positivity conditions}}

{\large These three features of presented here weak formulation are
chosen not for a provocative purpose or to illustrate a philosophical
doctrine that science is a matter of convention but to draw the attention
of practically oriented reader that they together can also be applied
in numerical methods. Linearity is associated with randomization or/and
smoothing of description, multitime is also related to a more complete
randomization description in which the time is not distinguish and
is treated as other variables. In result, the Kraichnan-Lewis multitime
correlation functions instead of the one time Reynolds' or Hopf's
correlation functions are used:}{\large \par}

{\large \begin{equation}
\varphi(\tilde{x})\Rightarrow<\varphi(\tilde{x}_{1})\cdots\varphi(\tilde{x}_{n})>;\; n=1,2,...,\infty\label{eq:3.1}\end{equation}
}{\large \par}

{\large Finally, the noncommuting variables are associated with an
additional generalization of the arena in which systems are described:
We introduce the free Fock space in which the correlation functions
or smoothed n-point functions (n-pfs) need not be permutation symmetric.
To stress this fact we will call n-pfs the }\textit{\large n-point
information}{\large . In this last step we do not project the original
equations, in our case equations for n-point information, but we enlarge
the space in which solutions are searched. This is exactly opposite
to what it is done in the Galerkin methods. Nevertheless, like in
the Galerkin methods, where corresponding space is diminished(!),
this permits the transfer of concepts of linear algebra to solve considered
equations. Among these concepts we take the generators for Cuntz algebra,
right and left invertible operators and plenty projectors using of
which allows us to construct varies final formulas for generating
vectors generating the correlation functions or, in the general case,
n-pfs, which we call the n-point information. In the case of the correlation
functions we have important restrictions:}{\large \par}

{\large \begin{equation}
<\varphi(x)^{2n}>\geq0\label{eq:3.2}\end{equation}
 for n=0,1,2...These restrictions we will call the }\textit{\large positivity
conditions, }{\large for correlation functions}\textit{\large . }{\large The
case}{\large \par}

{\large \begin{equation}
<\varphi(x)^{2n}>=0\Longleftrightarrow\varphi\equiv0\label{eq:3.3}\end{equation}
 means a trivial theory.}{\large \par}

\section{{\large The free Fock space and n-points information}}

{\large To deal with an infinite collection of correlation functions
or n-pfs, $<\varphi(\tilde{x}_{1})\cdots\varphi(\tilde{x}_{n})>$,
the one generating vector $|V>$can be introduced by means of which
all these n-pfs can be reproduced. Using such a vector we describe
the infinite system of branching equations for n-pfs in a compact
form of one vector equation which can be transformed in varies equivalent
and useful forms. From definition}{\large \par}

{\large \begin{eqnarray}
 & |V>=\nonumber \\
 & \sum_{n=1}\int d\tilde{x}_{(n)}<\varphi(\tilde{x}_{1})\cdots\varphi(\tilde{x}_{n})>\hat{\eta}^{\star}(\tilde{x}_{1})\cdots\hat{\eta}^{\star}(\tilde{x}_{n})|0>+V_{0}|0>\nonumber \\
\label{eq:4.1}\end{eqnarray}
where operators $\hat{\eta}^{\star}(\tilde{x})$, hermitian conjugat
to the operator $\hat{\eta}(\tilde{x}),$ satisfy the Cuntz relations}{\large \par}

{\large \begin{equation}
\hat{\eta}(\tilde{x})\hat{\eta}^{\star}(\tilde{y})=\hat{I}\cdot\delta(\tilde{x}-\tilde{y})\label{eq:4.2}\end{equation}
which mean that ranges of operators $\hat{\eta}^{\star}$ are pairwise
orthogonal and in fact the expantion (\ref{eq:4.1}) is a generalization
of the idea of expansion of a vector by means of a multiple orthogonal
base. Here operator $\hat{I}$- unit operator, $\delta$- is a product
of Kronecker's delta (discrete case) and vector |0> represents, using
quantum field theory language, a {}``vacuum'', }{\large \par}

{\large \begin{equation}
\hat{\eta}(\tilde{x})|0>=0\label{eq:4.3}\end{equation}
see\cite{han2007},\cite{han2008} and \cite{Hanckow (2010)}. Set
of vectors (\ref{eq:4.1}) form the free linear Fock space $F$. Lack
of commutation between quantities $\hat{\eta}*$in the generating
vectors |V> (free Fock space) does not exclude the possibility that
n-pfs $<\varphi(\tilde{x}_{1})\cdots\varphi(\tilde{x}_{n})>$are permutation
symmetric, see below. }{\large \par}

{\large Conditions (\ref{eq:4.2}) and (\ref{eq:4.3}) are enough
to show that }{\large \par}

{\large \begin{equation}
<0|\hat{\eta}(\tilde{y}_{1})\cdots\hat{\eta}(\tilde{y}_{n})|V>=<\varphi(\tilde{y}_{1})\cdots\varphi(\tilde{y}_{n})>\label{eq:4.4}\end{equation}
Since operators $\hat{\eta}$ do not commute, the above formula is
able to retrieve from the generating vector\ref{eq:4.1} also permutation
non-symmetrical n-pfs like in the case of quantum fields $\hat{\varphi}$.
But a true reason for introducing non-commuting field $\hat{\eta}$
is such that operators introduced below and constructed by means of
the operators $\hat{\eta,}\hat{\eta}^{\star}$ can be right or left
invertible, inverses to which can be easily constructed. This leads
to a variety of useful formulas for n-pfs. }{\large \par}

{\large We postulate the following equations for the n-pfs $<\varphi(\tilde{x}_{1})\cdots(\tilde{x}_{n})>$
which by means of the generating vector (\ref{eq:4.1}) can be described
in a compact way:}{\large \par}

{\large \begin{equation}
(\hat{L}+\lambda\hat{N}+\hat{G})|V>=\hat{P}_{0}|V>+\lambda\hat{P}_{0}\hat{N}|V>\equiv|0>_{ph}\label{eq:4.5}\end{equation}
with operators }{\large \par}

{\large \begin{eqnarray}
\hat{L}= & \int\hat{\eta}*(\tilde{x})L[\tilde{x};\hat{\eta}]d\tilde{x}+|0><0|=\nonumber \\
 & \int\hat{\eta}*(\tilde{x})L(\tilde{x},\tilde{y})\hat{\eta}(\tilde{y})d\tilde{x}d\tilde{y}+\hat{P}_{0}\nonumber \\
\label{eq:4.6}\end{eqnarray}
}{\large \par}

{\large \begin{equation}
\hat{N}=\int\hat{\eta}*(\tilde{z})N[\tilde{z};\hat{\eta}]d\tilde{z}+\hat{P}_{0}\hat{N}\label{eq:4.7}\end{equation}
 and }{\large \par}

{\large \begin{equation}
\hat{G}=\int\hat{\eta}*(\tilde{x})G(\tilde{x})\label{eq:4.8}\end{equation}
see\cite{han2007},\cite{han2008}. A small modification of the r.h.s.
of Eq.\ref{eq:4.5} is connected with a demand of right invertability
of the operators $\hat{L}$ and $\hat{N}$ what force us to add terms
$\hat{P}_{0}|V>$and $\lambda\hat{P}_{0}\hat{N}|V>$. For a concrete
choice of that term, see (\ref{eq:10.3'}). In fact, the $|0>_{ph}\neq|0>$
have to be used only for $\hat{G}\neq0$ and $\hat{N}\neq\hat{0}$.
It reminds us of a distant analogy with virtual particles of Quantum
Field Theory and therefore is called the }\textit{\large physical
vacuum. I}{\large n fact the r.h.s. of Eq.\ref{eq:4.5} comes from
the fact that the original Eq.\ref{eq:2.1} and the averaging process,
<...>, does not say anything about zero component of the equation.}{\large \par}

{\large Eq.\ref{eq:4.5} means that we have chosen averages with respect
to used additional conditions (ensemble averages). As we said in Sec.3,
in the case of homogeneous system, both types of averages lead to
the same equations for the correlation functions. }{\large \par}

{\large By introducing projectors $\hat{P}_{n}$ projecting on the
consecutive terms of the expansion (\ref{eq:4.1}), we can express
the projection properties of operators (\ref{eq:4.6}-\ref{eq:4.8})
as follows:}{\large \par}

{\large \begin{equation}
\hat{P}_{n}\hat{L}=\hat{L}\hat{P}_{n}\label{eq:4.9}\end{equation}
(diagonal), where n=0,1,2,...,}{\large \par}

{\large \begin{equation}
\hat{P}_{n}\hat{N}=\sum_{n<m}\hat{P}_{n}\hat{N}\hat{P}_{m}\label{eq:4.10}\end{equation}
(upper triangular), where n=0,1,2,..., see (\ref{eq:4.7}) and (\ref{eq:10.3'}). }{\large \par}

{\large \begin{equation}
\hat{P}_{n}\hat{G}=\hat{G}\hat{P}_{n-1}\label{eq:4.11}\end{equation}
(lower triangular), where n=1,2,.... The operator values function
$N[\tilde{z};\hat{\eta}]$ can be a polynomial or other function depending
on the vector variable $\tilde{z}$ and the operator variables $\hat{\eta}(\tilde{x})$
indexed by the vector variable $\tilde{x}$. The operator $\hat{N}$
is related to a nonlinear part of the strong formulation of theory
(the original differential equations (\ref{eq:2.1}). The operator
$\hat{G}$ describes a source term with a function $G(\tilde{x})$
correponding to the external forces, for example. It is symtomatic
that diagonal and upper triangular operators describe an interaction
or selfinteraction of the constituents of the system and that lower
triangular operators describe an interaction with the external world
or quantum properties of the system (microworld). }{\large \par}

{\large The simplest diagonal operator is the unit operator}{\large \par}

{\large \begin{equation}
\hat{I}=|0><0|+\int\hat{\eta}*(\tilde{x})\hat{\eta}(\tilde{x})d\tilde{x}\label{eq:4.12}\end{equation}
Other diagonal operators are the projectors used in formulas (\ref{eq:4.9}-\ref{eq:4.11})
and constructed by means of the tensor product of vectors:}{\large \par}

{\large \begin{eqnarray}
 & \hat{P}_{n}=\int\hat{\eta}*(\tilde{x}_{1})\cdots\hat{\eta}*(\tilde{x}_{n})|0><0|\hat{\eta}(\tilde{x}_{n})\cdots\hat{\eta}(\tilde{x}_{1})d\tilde{x}_{(n)}\nonumber \\
\label{eq:4.13}\end{eqnarray}
 where $\hat{P}_{0}=|0><0|$. They form a complete set of orthogonal
projectors:}{\large \par}

{\large \begin{equation}
\sum_{n=0}\hat{P}_{n}=\hat{I},\quad and\;\hat{P}_{m}\hat{P}_{n}=\hat{P}_{n}\delta_{mn}\label{eq:4.14}\end{equation}
 We can say that projections $\hat{P}_{n}|V>$, for n=1,2,..., provide
n-points information about the local nature of the system but the
projection $\hat{P}_{0}|V>$provides rather global, agregated information.
In the case of system representing the Universe, the r.hs. of Eqs
like (\ref{eq:4.5}), (\ref{eq:7.1}), (\ref{eq:7.11}), (\ref{eq:7.15})
can be interpreted as a vacuum, see \cite{Hanckow (2010)}. Thus,
in this interpretation the (classical) vacuum contains the global
information about the Universe. Like in QFT a non-trivial structure
of the vacuum arises only through the nonlinear theory and this is
a positive element that might enable its stady. Identical equations
as (\ref{eq:4.5}) take place in QFT, for vacuum expectation values}{\large \par}

\textit{\large \begin{equation}
<\hat{\varphi}(\tilde{x}_{1})...\hat{\varphi}(\tilde{x}_{n})>\label{eq:4,15}\end{equation}
}{\large where $\hat{\varphi}$are now operators with appropriate
equal time commutators. In this case however, n-pfs (\ref{eq:4.4})
are not permutationally symmetric.}{\large \par}

\section{{\large Right invertible linear part of the theory and approximated
solutions }}

{\large If the kernel $L$ of a diagonal operator $\hat{L}$ is a
right invertible:}{\large \par}

{\large \begin{equation}
\int L(\tilde{x},\tilde{y})L_{R}^{-1}(\tilde{y},\tilde{z})d\tilde{z}=\delta(\tilde{x}-\tilde{z})\label{eq:5.1}\end{equation}
then the operator $\hat{L}$ is a right invertible in the Fock space
$F$, see \ref{eq:4.1}-\ref{eq:4.3}. A right inverse to $\hat{L}$,
denoted by $\hat{L}_{R}^{-1}$can be constructed as }{\large \par}

{\large \begin{equation}
\hat{L}_{R}^{-1}=\int\hat{\eta}*(\tilde{z})L_{R}^{-1}(\tilde{z},\tilde{w})\hat{\eta}(\tilde{w})d\tilde{z}d\tilde{w}+\hat{P}_{0}\label{eq:5.2}\end{equation}
where, as before, the symbol $\int$ means the sumation or integration
with respect to components of vectors $\tilde{z,}\tilde{w}$. In fact,
the operator $\hat{L}_{R}^{-1}$satisfies the weaker equation: }{\large \par}

{\large \begin{equation}
\hat{L}\hat{L}_{R}^{-1}=\hat{I}\label{eq:5.3}\end{equation}
}{\large \par}

{\large Having constructed a right inverse to a given operator $\hat{L}$,
we can construct the projector on the null space of $\hat{L}$:}{\large \par}

{\large \begin{equation}
\Pi_{L}=\hat{I}-\hat{L}_{R}^{-1}\hat{L}\label{eq:5.4}\end{equation}
 Multiplying Eq.\ref{eq:4.5} by a right inverse operator $(\hat{L}+\hat{G})$,
we can describe this equation in an equivalent way as follows}{\large \par}

{\large \begin{eqnarray}
 & [\hat{I}+\lambda(\hat{L}+\hat{G})_{R}^{-1}\hat{N}]|V>=\hat{\Pi}_{L+G}|V>+\nonumber \\
 & (\hat{L}+\hat{G})_{R}^{-1}\left(\hat{P}_{0}|V>+\lambda\hat{P}_{0}\hat{N}|V>\right)\nonumber \\
\label{eq:5.5}\end{eqnarray}
where the projector on the null space of the operator $(\hat{L}+\hat{G})$
is}{\large \par}

{\large \begin{equation}
\hat{\Pi}_{L+G}=\hat{I}-(\hat{L}+\hat{G})_{R}^{-1}(\hat{L}+\hat{G})\label{eq:5.5'}\end{equation}
}{\large \par}

{\large From point of view of Eq.\ref{eq:4.5}, considered in the
free (full, super) Fock space $F$, the projection $\hat{\Pi}_{L+G}|V>$can
be any vector from space $\hat{\Pi}_{L+G}F$. With different right
inverse operators$\hat{L}_{R}^{-1}$, different projectors $\hat{\Pi}_{L+G}$on
the null space of the operator $(\hat{L}+\hat{G})$ are constructed.
In result, a general solution to the generating vector |V> obtained
by means Eq.\ref{eq:5.5} has different parametrizations. }{\large \par}

{\large From definition of smoothing operations given in Sec.3, we
see that obtained n-pfs are permutation symmetric. Denoting by $\hat{S}$
a projector on vectors generating permutation symmetric n-pfs , we
should have, for the }\textit{\large physical solutions}{\large :}{\large \par}

{\large \begin{equation}
|V>=\hat{S}|V>\label{eq:5.6}\end{equation}
 Hence and from Eq.\ref{eq:5.5} , we get }{\large \par}

{\large \begin{eqnarray}
 & [\hat{I}+\lambda\hat{S}(\hat{L}+\hat{G})_{R}^{-1}\hat{N}]|V>=\nonumber \\
 & \hat{S}\hat{\Pi}_{L+G}|V>+\hat{S}(\hat{L}+\hat{G})_{R}^{-1}\left(\hat{P}_{0}|V>+\lambda\hat{P}_{0}\hat{N}|V>\right)\nonumber \\
\label{eq:5.7}\end{eqnarray}
Eq.\ref{eq:5.7} was derived by using first Eq.\ref{eq:5.6} and next
acting on Eq.\ref{eq:5.5} with projector $\hat{S}$. Similar procedure
is used in the Galerkin method, however, there is an important difference
with Galerkin approach, namely - Eq.\ref{eq:5.6} - is an exact equation
and hence the solutions to Eq.\ref{eq:5.7} can also be exact solutions
of the original problem.}{\large \par}

{\large Taking into account the projection properties of operators
$\hat{S},\hat{L}_{R}^{-1}$ and $\hat{G}$, Eq.\ref{eq:4.5} can be
also described as: }{\large \par}

{\large \begin{eqnarray}
 & \{\hat{I}+\lambda(\hat{I}+\hat{S}\hat{L}_{R}^{-1}\hat{G})^{-1}\hat{S}\hat{L}_{R}^{-1}\hat{N}\}|V>=\nonumber \\
 & \hat{S}\hat{\Pi}_{L}|V>+\hat{S}(\hat{L}+\hat{G})_{R}^{-1}\left(\hat{P}_{0}|V>+\lambda\hat{P}_{0}\hat{N}|V>\right)\nonumber \\
\label{eq:5.8}\end{eqnarray}
 }{\large \par}

{\large This or Eq.\ref{eq:5.7} are the output equatios for calculating
successive approximations to the generating vector |V> expanded in
the positive powers of the coupling $\lambda$ standing at the nonlinear
part of original theory (\ref{eq:2.1}):}{\large \par}

{\large \begin{equation}
|V>=\sum_{j=0}^{\propto}\lambda^{j}|V>^{(j)}\label{eq:5.9}\end{equation}
 The arbitary element of Eq.\ref{eq:5.5}, $\hat{\Pi}_{L}|V>\in\hat{\Pi}_{L}F$,
can also be expanded in this way }{\large \par}

{\large \begin{equation}
\hat{\Pi}_{L}|V>=\sum_{j=0}^{\propto}\lambda^{j}(\hat{\Pi}_{L}|V>)^{(j)}\label{eq:5.10}\end{equation}
 The terms of the above series can be restricted by the permutation
symmetry of n-pfs. In the case of symmetrical solutions (\ref{eq:5.6})
to Eq.\ref{eq:5.7} we can assume that the symmetry part of the arbitrary
element $\hat{\Pi}_{L}|V>$is given by a linear theory:}{\large \par}

{\large \begin{equation}
\hat{S}\hat{\Pi}_{L+G}|V>=\hat{S}|V>^{(0)}=|V>^{(0)}\label{eq:5.10'}\end{equation}
It is a common situation accompaning to almost every approximated
and exact theory, see also Sec.8. It means that only due to terms
with $\lambda\neq0$ higher approximations appear. So the zeroth order
approximation}{\large \par}

{\large \begin{equation}
|V>^{(0)}=\hat{S}\hat{\Pi}_{L}|V>^{(0)}\label{eq:5.11}\end{equation}
 The first order approximation}{\large \par}

{\large \begin{equation}
|V>^{(1)}=-\{(\hat{I}+\hat{S}\hat{L}_{R}^{-1}\hat{G})^{-1}\hat{S}\hat{L}_{R}^{-1}\hat{N}\}|V>^{(0)}\label{eq:5.12}\end{equation}
 The second order approximation}{\large \par}

{\large \begin{equation}
|V>^{(2)}=-\{(\hat{I}+\hat{S}\hat{L}_{R}^{-1}\hat{G})^{-1}\hat{S}\hat{L}_{R}^{-1}\hat{N}\}|V>^{(1)}\label{eq:5.13}\end{equation}
and so on. All these approximations can be obtained from a single
vector formula: }{\large \par}

{\large \begin{eqnarray}
 & |V>=\nonumber \\
 & \{\hat{I}+\lambda(\hat{I}+\hat{S}\hat{L}_{R}^{-1}\hat{G})^{-1}\hat{S}\hat{L}_{R}^{-1}\hat{N}\}^{-1}(\hat{I}+\hat{S}\hat{L}_{R}^{-1}\hat{G})^{-1}\hat{S}\hat{\Pi}_{L}|V>\nonumber \\
\label{eq:5.13'}\end{eqnarray}
 and it is not inconceivable that these compact formula provides a
new look at old divergent problems of unrenormalizable theories. }{\large \par}

{\large Let us focuse on the zeroth order approximation, (\ref{eq:5.11}).
It is a symmetrical solution to the Eq.\ref{eq:4.5}, for $\lambda=0$.
We will consider this equation with additional simplification that
exterial forces acting on the system and represented by the operator
$\hat{G}=0$. So we get the following vector equation }{\large \par}

{\large \begin{equation}
\hat{L}|V>^{(0)}=0\label{eq:5.14}\end{equation}
 describing a linear theory. In the {}``components'' form it looks
as follows:}{\large \par}

{\large \begin{equation}
\int d\tilde{x}_{1}L(\tilde{x},\tilde{x}_{1})<\varphi(\tilde{x}_{1})\cdots\varphi(\tilde{x}_{n})>^{(0)}=0\label{eq:5.15}\end{equation}
 for n=1,2,... Let us assume that we know a function $\triangle$
which satisfies equation}{\large \par}

{\large \begin{equation}
\int d\tilde{x}_{1}L(\tilde{x},\tilde{x}_{1})\triangle(\tilde{x}_{1},\tilde{y})=0\label{eq:5.16}\end{equation}
 for all $\tilde{x},\tilde{y}$. With functions $\triangle$, we can
construct, for even n, symmetrical solutions to Eqs \ref{eq:5.15}
as follows:}{\large \par}

{\large \begin{eqnarray}
 & <\varphi(\tilde{x}_{1})\cdots\varphi(\tilde{x}_{n})>^{(0)}\equiv V(\tilde{x}_{1},...,\tilde{x}_{n})^{(0)}=\nonumber \\
 & \frac{1}{n!}\sum_{perm}\triangle(\tilde{x}_{1},\tilde{x}_{2})\triangle(\tilde{x}_{3},\tilde{x}_{4})\cdots\triangle(\tilde{x}_{n-1},\tilde{x}_{n})\nonumber \\
\label{eq:5.17}\end{eqnarray}
and zero, for odd n. Because of symmetry, the permutation operation
is taken in the sum. In a graph representation, with vertexes denoted
by $\tilde{x}_{1},...,\tilde{x}_{n}$, a particular terms of sum \ref{eq:5.17}are
represented by graphs with one edge vertexes. It is easy to see that,
for such solutions, conditions (\ref{eq:3.2}) are satisfied. In addition,
we also expect that these conditions are also fulfilled for the higher
order approximations for the correlation functions when the perturbation
parameter, the coupling constant $\lambda$, is a small quantity in
Eq.\ref{eq:5.8}. }{\large \par}

{\large As an example of \ref{eq:5.17},}{\large \par}

{\large \begin{equation}
<\varphi(\tilde{x})>^{(0)}=0\label{eq:5.18}\end{equation}
 \begin{equation}
<\varphi(\tilde{x})\varphi(\tilde{y})>^{(0)}=\frac{1}{2!}(\triangle(\tilde{x},\tilde{y})+\triangle(\tilde{y},\tilde{x}))=\triangle(\tilde{x},\tilde{y})\label{eq:5.19}\end{equation}
 where the permutation symmetry of function $\triangle$ is assumed.
Hence we see that in zeroth order approximation the correlation functions
(\ref{eq:5.17}) are sums of products of the 2-point function\ref{eq:5.19}. }{\large \par}

{\large To find a connection of the zeroth order 2-pf $\Delta$ with
the general solution to Eq.\ref{eq:2.1}, in the case of $\lambda=0,\, G=0$,
we represent this solution in the form }{\large \par}

{\large \begin{equation}
\varphi^{(0)}[\tilde{x},\alpha]=\int d\tilde{u}\Gamma(\tilde{x},\tilde{u})\alpha(\tilde{u})\label{eq:5.20}\end{equation}
in which the }\textit{\large kernel}{\large{} $\Gamma$ of the general
solution (\ref{eq:2.1}) ($\lambda=0,G=0$) represents possible elementary
solutions to Eqs (\ref{eq:2.1}) and\ref{eq:5.16} - labeled by the
vector variable $\tilde{u}$ which are summed with $\tilde{u}$- dependent
factor $\alpha$. We do not define here a set to which the vector
parameteru $\tilde{u}$ belongs. From \ref{eq:5.19}) and \ref{eq:5.20}
we get }{\large \par}

{\large \begin{eqnarray}
 & <\varphi(\tilde{x})\varphi(\tilde{y})>^{(0)}\equiv\triangle(\tilde{x},\tilde{y})=\nonumber \\
 & \int d\tilde{u}d\tilde{w}\delta\alpha\Gamma(\tilde{x},\tilde{u})\Gamma(\tilde{y},\tilde{w})\alpha(\tilde{u})\alpha(\tilde{w})P[\alpha]\nonumber \\
\label{eq:5.21}\end{eqnarray}
 Denoting the functional integral occurring in (\ref{eq:5.21}) as }{\large \par}

{\large \begin{equation}
\int\delta\alpha\alpha(\tilde{u})\alpha(\tilde{w})P[\alpha]=P(\tilde{u},\tilde{w})\label{eq:5.22}\end{equation}
 we get for the function $\triangle$, the following expression: }{\large \par}

{\large \begin{eqnarray}
 & \triangle(\tilde{x},\tilde{y}) & =<\varphi(\tilde{x})\varphi(\tilde{y})>^{(0)}=\int d\tilde{u}d\tilde{w}\Gamma(\tilde{x},\tilde{u})\Gamma(\tilde{y},\tilde{w})P(\tilde{u,}\tilde{w})\nonumber \\
\label{eq:5.23}\end{eqnarray}
which relates the correlation function $\triangle$ between cells
with the }\textit{\large kernel}{\large{} $\Gamma$ of the general
solution to (\ref{eq:2.1}), at $\lambda=0,$G=0, and the }\textit{\large second
order moments}{\large{} $P(\tilde{u},\tilde{w})$ of the }\textit{\large probability
density}{\large{} $P[\alpha]$. }{\large \par}

{\large The kernel $\Gamma$ of the general solution to Eq.\ref{eq:2.1},
in case ($\lambda=0,\, G=0$), can be constructed by means of a right
inverse $L_{R}^{-1}$, see \ref{eq:5.1}: }{\large \par}

{\large \begin{equation}
\Gamma(\tilde{x},\tilde{u})\equiv\Pi_{L}(\tilde{x},\tilde{u})=\delta(\tilde{x}-\tilde{u})-\int d\tilde{z}L_{R}^{-1}(\tilde{x},\tilde{z})L(\tilde{z},\tilde{u})\label{eq:5.24}\end{equation}
 In such case this is a projector (idempotent): }{\large \par}

{\large \begin{equation}
\Gamma(\tilde{x},\tilde{u})=\int d\tilde{z}\Gamma(\tilde{x},\tilde{z})\Gamma(\tilde{z},\tilde{u})\label{eq:5.25}\end{equation}
and the subspace described by the projector $\Gamma$ can be identified
with all solutions to Eq.\ref{eq:2.1} . In many cases, with every
subspce, one can define a linear differential equation, see \cite{Svierish (1996)}.
In fact a differential equation can be }\textit{\large represented
geometrically}{\large{} by the set of all solutions, and a }\textit{\large symmetry
of the differential equation}{\large{} is defined as a map which transforms
this set into itself,\cite{Hydon (2005)}. }{\large \par}

{\large For a symmetrical $\Gamma$ }{\large \par}

{\large \begin{equation}
\Gamma(\tilde{x},\tilde{y})=\Gamma(\tilde{y},\tilde{x})\label{eq:5.26}\end{equation}
and the Gaussian}{\large \par}

{\large \begin{equation}
P(\tilde{u},\tilde{w})=\delta(\tilde{u}-\tilde{w})\label{eq:5.27}\end{equation}
we get from (\ref{eq:5.21}) }{\large \par}

{\large \begin{equation}
\triangle(\tilde{x},\tilde{y})\equiv<\varphi(\tilde{x})\varphi(\tilde{y})>^{(0)}=\Gamma(\tilde{x},\tilde{y})\label{eq:5.28}\end{equation}
 In other words, for the linear systems (\ref{eq:2.1}) described
by the symmetrical, idempotent kernel$\Gamma$of the general solution
and the Gaussian type moments (\ref{eq:5.27}), the zeroth order 2-pf
is identical with this kernel (sic!). It is astonishing case which
shows that, at least at the zeroth order level, information lost by
averaging or smoothing procedures ($<\varphi(\tilde{x})>^{(0)}=0\, for\, all\,\tilde{x}$)
can be recovered with the help of 2-point correlation function (\ref{eq:5.28}).
It is not excluded that the following hypothesis is true: - in the
case of nonlinear equations a similar phenomen takes place with the
help of the higher order correlation functions and the Gauss smoothing. }{\large \par}

{\large For $\varphi^{3}$nonlinear, local theory, the above properties
are particularly important because they allow to substitute the bilinear
products $\Delta\cdot\Delta$occurring in such a theory by the one
$\Delta$, see Sec.5. }{\large \par}

{\large It is interesting to notice that condition (\ref{eq:5.26})
called sometimes the }\textit{\large reverse normalized condition
}{\large is in fact one additional condition imposed on a right inverse
operator to get a solution to (\ref{eq:2.1}) with smallest Euclidean
norm (generalized inverse or pseudo-inverse called also Moore-Penrose
inverse). This illustrates the kind of processes described by Eq.\ref{eq:5.28}.
It turns out that every bounded operator $L:\, K\rightarrow H$ (Hilbert
spaces) with closed range has a generalized inverse. To see properties
of such operators, see\cite{Christensen (1999)}. }{\large \par}

\section{{\large Right invertible {}``nonlinear'' part of the theory }}

{\large In this case we will assume and it actually takes place in
the free Fock space, \cite{Hanckow (2007)}-\cite{Hanckow (2010)}
that a right inverse operator $\hat{N}_{R}^{-1}$exists to the operator
$\hat{N}$:}{\large \par}

{\large \begin{equation}
\hat{N}\hat{N}_{R}^{-1}=\hat{I}\label{eq:6.1}\end{equation}
and that the operator $\hat{N}_{R}^{-1}$can be explicitly constructed
in the most general form, \cite{Hanckow (2007)}-\cite{Hanckow (2010)}.
Introducing a projector on the null space of the operator $\hat{N}$: }{\large \par}

{\large \begin{equation}
\hat{P}=\hat{I}-\hat{N}_{R}^{-1}\hat{N}\equiv\hat{I}-\hat{Q}_{N};\quad<0|\hat{N}=0\label{eq:6.2}\end{equation}
 we can write equivalently Eq.\ref{eq:4.5} as follows:}{\large \par}

{\large \begin{eqnarray}
 & [\hat{I}+\lambda^{-1}\hat{N}_{R}^{-1}(\hat{L}+\hat{G})]|V>=\hat{P}|V>+\nonumber \\
 & \lambda^{-1}\hat{N}_{R}^{-1}\left(\hat{P}_{0}|V>+\lambda\hat{P}_{0}\hat{N}|V>\right)\label{eq:6.3}\end{eqnarray}
 With symmetry (\ref{eq:5.6}) we get}{\large \par}

{\large \begin{eqnarray}
 & [\hat{I}+\lambda^{-1}\hat{S}\hat{N}_{R}^{-1}(\hat{L}+\hat{G})]|V>=\hat{S}\hat{P}|V>+\nonumber \\
 & \lambda^{-1}\hat{S}\hat{N}_{R}^{-1}\left(\hat{P}_{0}|V>+\lambda\hat{P}_{0}\hat{N}|V>\right)\label{eq:6.4}\end{eqnarray}
} {\large Assuming that with limes $\lambda\rightarrow0$ we get finite
results for the generating vector |V>, we have}{\large \par}

\begin{equation}
|V(\lambda=\infty)>=\hat{S}\hat{P}|V(\infty)>+\hat{P}_{0}\hat{N}|V(\infty)>\label{eq:6.5}\end{equation}
{\large{} In a spirit of perturbation theory we assume that the arbitrary
term of solutions to Eq.\ref{eq:6.4} }{\large \par}

{\large \begin{equation}
\hat{S}\hat{P}|V>=\hat{S}\hat{P}|V(\infty)>=|V(\lambda=\infty)>-\hat{P}_{0}\hat{N}|V(\infty)>=?\label{eq:6.6}\end{equation}
 and a possible answer to that is a real issue.}{\large \par}

\section{{\large Modified or regularized theory}}

{\large We mean by this an introducing to the term representing the
nonlinear part of the original theory - one extra parameter: $\hat{N}\rightarrow\hat{N}(\lambda_{2})$.
This parameter is called the }\textit{\large minor coupling constant
}{\large and is denoted by}\textit{\large{} $\lambda_{2}$}{\large{}
where the original parameter $\lambda$, now denoted by $\lambda_{1}$,
will be called the }\textit{\large major coupling constant. }{\large If
at the end of calculations we go with parameter $\lambda_{2}$ to
zero and with $\lambda_{2}\rightarrow0$, $\hat{N}(\lambda_{2})\rightarrow\hat{N}$
we will speak about a regularization of the theory in other case about
its modification. In certain cases such theories with two coupling
constants $\lambda_{1},\lambda_{2}\neq0$ satisfies the Dyson criterion
of convergency,\cite{Salam et al. (1971)}, and hence the name -}\textit{\large{}
regularization}{\large . In other words, divergences of the usual
perturbation theory can be interpreted as residues of the deformed,
nonlocal and more basic theory which is currently still unknown. For
other interesting ideas concerning renormalizability of a theory see
\cite{Kosyakov (2000)}. }{\large \par}

{\large So, instead of Eq.\ref{eq:4.5} we consider the }\textit{\large resolvent
regularized or modified equation}{\large{} }{\large \par}

{\large \begin{eqnarray}
 & \left(\hat{L}+\lambda_{1}(\hat{I}+\lambda_{2}\hat{M})_{R}^{-1}\hat{N}+\hat{G}\right)|V>=\nonumber \\
 & \hat{P}_{0}|V>+\lambda_{1}\hat{P}_{0}(\hat{I}+\lambda_{2}\hat{M})_{R}^{-1}\hat{N}|V>\equiv|0>_{reg}\label{eq:7.1}\end{eqnarray}
where the operator $(\hat{I}+\lambda_{2}\hat{M})_{R}^{-1}$is a right
inverse operator to the operator $(\hat{I}+\lambda\hat{M})$. In the
mathematical literature, this operator is called a }\textit{\large generalized
resolvent of the operator $\hat{M}$,}{\large{} \cite{Kato (1966)}.
See \cite{Ivan (1974)}, for other use of expression - the resolvent
equations.}{\large \par}

{\large The r.h.s. of Eq.\ref{eq:7.1} comes from the fact that the
original Eq.\ref{eq:2.1} and the averaging process, <...>, does not
say anything about zero component of the equation. Therefore, acting
on this equation with projector $\hat{P}_{0}$we should get identity. }{\large \par}

{\large Multiplying the last equation by the operator $(\hat{I}+\lambda\hat{M})$,
we remove, at least from the l.h.s. of Eq.\ref{eq:7.1}, ambiguities
contained into the right inverse operator and get }{\large \par}

{\large \begin{eqnarray}
 & \left[(\hat{I}+\lambda_{2}\hat{M})(\hat{L}+\hat{G})+\lambda_{1}\hat{N}\right]|V>=\nonumber \\
 & (\hat{I}+\lambda_{2}\hat{M})\left(\hat{P}_{0}|V>+\lambda_{1}\hat{P}_{0}(\hat{I}+\lambda_{2}\hat{M})_{R}^{-1}\hat{N}|V>\right)\label{eq:7.2}\end{eqnarray}
This equation can also be treated as a }\textit{\large regularization}{\large{}
of the original Eq.\ref{eq:4.5} in a sense that, for $\lambda_{2}\rightarrow0$,
(\ref{eq:7.2}) tends to Eq.\ref{eq:4.5}. }{\large \par}

{\large Eq.\ref{eq:7.2} is not equivalent to (\ref{eq:7.1}\} because
it was derived by multiplying by the right invertible operator $(\hat{I}-\lambda_{2}\hat{M})$.
This operation is equivalent to the operation of multiplication by
the projector }{\large \par}

{\large \begin{equation}
\hat{Q}_{(I+\lambda_{2}M)}=(\hat{I}+\lambda_{2}\hat{M})_{R}^{-1}(\hat{I}+\lambda_{2}\hat{M})\label{eq:7.3}\end{equation}
 We have here an interesting situation in which the projected Eq.\ref{eq:7.2}
tends in the limes, $\lambda_{2}\rightarrow0$, to the original Eq.\ref{eq:4.5}.
Is it possible from the point of view of logic? The answer to this
question is positive, if we take into account that equation (4.5)
considered in the Fock space are }\textit{\large overdetermined}{\large .
Why? Because Eqs$\ref{eq:4.5}$ or its regularized version (\ref{eq:7.1})
or (\ref{eq:7.2}) are exactly the same in the free Fock space as
well as in the physical space (\ref{eq:5.6}). }{\large \par}

{\large Now we transform the Eq.\ref{eq:7.2} in an equivalent way
taking into account the right invertibility of the operators $\hat{M}$
and $\hat{L}$. We get}{\large \par}

{\large \begin{eqnarray}
 & \left\{ \hat{I}+[\lambda_{2}\hat{M}(\hat{L}+\hat{G})]_{R}^{-1}[\hat{L}+\hat{G}+\lambda_{1}\hat{N}]\right\} |V>=\hat{P}_{M}|V>+\nonumber \\
 & [\lambda_{2}\hat{M}(\hat{L}+\hat{G})]_{R}^{-1}\left(\hat{P}_{0}|V>+\lambda_{1}\hat{P}_{0}(\hat{I}+\lambda_{2}\hat{M})_{R}^{-1}\hat{N}|V>\right)\nonumber \\
\label{eq:7.4}\end{eqnarray}
where }{\large \par}

{\large \begin{equation}
\hat{P}_{M}=\hat{I}-\hat{[M}(\hat{L}+\hat{G})]_{R}^{-1}\hat{M}(\hat{L}+\hat{G})\label{eq:7.5}\end{equation}
is a projector on the null space of the operator $\hat{M}(\hat{L}+\hat{G})$.
This projector has the following structure:}{\large \par}

{\large \begin{equation}
\hat{P}_{M}=\hat{I}-(\hat{L}+\hat{G})_{R}^{-1}\hat{Q}_{M}(\hat{L}+\hat{G})\label{eq:7.6}\end{equation}
with projector $\hat{Q}_{M}=\hat{M}_{R}^{-1}\hat{M}$ and a right
inverse }{\large \par}

{\large \begin{equation}
(\hat{L}+\hat{G})_{R}^{-1}=\hat{L}_{R}^{-1}(\hat{I}+\hat{G}\hat{L}_{R}^{-1})_{R}^{-1}\label{eq:7.7}\end{equation}
}{\large \par}

{\large The projected vector of |V>, $\hat{P}_{M}|V>$, is an arbitrary
vector of the null space of the operator $\hat{M}(\hat{L}+\hat{G})$.
It describes an additional freedom which a deformed theory brings
to the original theory given by Eq.\ref{eq:4.5}. From (\ref{eq:5.5'})
we have}{\large \par}

{\large \begin{equation}
\hat{P}_{M}\hat{\Pi}_{L+G}=\hat{\Pi}_{L+G}\label{eq:7.7'}\end{equation}
}{\large \par}

{\large It is also possible to have a different perspective on the
Eq.\ref{eq:7.1}. This time we will not weaken this equation but we
represent the operator $\hat{R}(\lambda_{2})\equiv(\hat{I}+\lambda_{2}\hat{M})_{R}^{-1}$
in a more explicit way assuming that $\hat{M}$ is a right invertible
operator. Then we get }{\large \par}

{\large \begin{eqnarray}
 & \hat{R}(\lambda_{2})\equiv(\hat{I}+\lambda_{2}\hat{M})_{R}^{-1}=\nonumber \\
 & [\hat{I}+(\lambda_{2}\hat{M})_{R}^{-1}]^{-1}[(\lambda_{2}\hat{M})_{R}^{-1}+\hat{\Gamma}_{M}\hat{R}(\lambda_{2})]\label{eq:7.8}\end{eqnarray}
 with an arbitrary element $\hat{\Gamma}_{M}\hat{R}(\lambda_{2})$
where the projector }{\large \par}

{\large \begin{equation}
\hat{\Gamma}_{M}=\hat{I}-\hat{M}_{R}^{-1}\hat{M}\equiv\hat{I}-\hat{Q}_{M}\label{eq:7.9}\end{equation}
It is interesting that a choice}{\large \par}

{\large \begin{equation}
\hat{\Gamma}_{M}\hat{R}(\lambda_{2})=0\label{eq:7.10}\end{equation}
leads to a generalized resolvent $\hat{R}(\lambda_{2})$ satisfying
the resolvent equation as}{\large \par}

{\large \begin{equation}
\frac{d}{d\lambda_{2}}\hat{R}=-\hat{M}\hat{R}^{2}\label{eq:7.100}\end{equation}
 which is also satisfied for the operators $\hat{M}$ with the usual
resolvents. This equation, in turn, ensures that, for at least right
invertible operators $\hat{M}$, the choice $\hat{R}=[\hat{I}+(\lambda_{2}\hat{M})_{R}^{-1}]^{-1}(\lambda_{2}\hat{M})_{R}^{-1}$
satisfies the same condition as for the non singular $\hat{M}$, namely:
$\hat{R}\rightarrow0$, for $\lambda_{2}\rightarrow\infty$. In practical
considerations, the term $\hat{\Gamma}_{M}\hat{R}(\lambda_{2})\neq0$
can be used when there are problems with limes $\lambda_{2}\rightarrow0$. }{\large \par}

{\large Let us also notice that in the zero-$\lambda_{2}limit$ the
vector }{\large \par}

{\large \begin{equation}
|\Psi(\lambda_{2})>\equiv[\hat{I}+(\lambda_{2}\hat{M})_{R}^{-1}]^{-1}(\lambda_{2}\hat{M})_{R}^{-1}|\Phi>\rightarrow|\Phi>\label{eq:7.10'}\end{equation}
 at the assumption}{\large \par}

{\large \begin{equation}
\lambda_{2}\hat{M}|\Psi(\lambda_{2})>\rightarrow0,\quad for\;\lambda_{2}\rightarrow0\label{eq:7.10''}\end{equation}
 because from (\ref{eq:7.10'}) results that}{\large \par}

{\large \begin{equation}
[\lambda_{2}\hat{I}+\hat{M}{}_{R}^{-1}]|\Psi(\lambda_{2})>=\hat{M}{}_{R}^{-1}|\Phi>\label{eq:7.10'''}\end{equation}
and finally}{\large \par}

{\large \begin{equation}
[\lambda_{2}\hat{M}+\hat{I}]|\Psi(\lambda_{2})>=|\Phi>\label{eq:7.10''''}\end{equation}
 In other words the operator }{\large \par}

{\large \begin{equation}
[\hat{I}+(\lambda_{2}\hat{M})_{R}^{-1}]^{-1}(\lambda_{2}\hat{M})_{R}^{-1}\rightarrow\hat{I}\label{eq:7.10'''''}\end{equation}
 behaves as the unit operator, with $\lambda_{2}\rightarrow0$, if
the vector $\hat{M}|\Psi(\lambda_{2})>$has oscillations suppressed
by $\lambda_{2}\rightarrow0$. In fact, the paper presented is about
the indeterminate expression $0\cdot\infty$ or rather -$\hat{0}\cdot\hat{\infty}$-
because we are dealing here with operators. }{\large \par}

{\large Eq.\ref{eq:7.1} with representation (\ref{eq:7.8}) is following:}{\large \par}

{\large \begin{eqnarray}
 & \left(\hat{L}+\hat{G}\right)|V>+\nonumber \\
 & \lambda_{1}\left([\hat{I}+(\lambda_{2}\hat{M})_{R}^{-1}]^{-1}[(\lambda_{2}\hat{M})_{R}^{-1}+\hat{\Gamma}_{M}\hat{R}(\lambda_{2})]\right)\hat{N}|V>\nonumber \\
 & =\hat{P}_{0}|V>+\lambda_{1}\hat{P}_{0}(\hat{I}+\lambda_{2}\hat{M})_{R}^{-1}\hat{N}|V>\label{eq:7.11}\end{eqnarray}
 With a right inverse operator, $(\hat{L}+\hat{G})_{R}^{-1}$, we
can transform this equation in an equivalent way as}{\large \par}

{\large \begin{eqnarray}
 & |V>+\lambda_{1}(\hat{L}+\hat{G})_{R}^{-1}[\hat{I}+(\lambda_{2}\hat{M})_{R}^{-1}]^{-1}\cdot\nonumber \\
 & \left([(\lambda_{2}\hat{M})_{R}^{-1}+\hat{\Gamma}_{M}\hat{R}(\lambda_{2})]\right)\hat{N}|V>\nonumber \\
 & =\hat{\Pi}_{L+G}|V>+\nonumber \\
 & (\hat{L}+\hat{G})_{R}^{-1}\left(\hat{P}_{0}|V>+\lambda_{1}\hat{P}_{0}(\hat{I}+\lambda_{2}\hat{M})_{R}^{-1}\hat{N}|V>\right)\label{eq:7.12}\end{eqnarray}
 where a projector on the null space of the operator $(\hat{L}+\hat{G})$
is given by }{\large \par}

{\large \begin{equation}
\hat{\Pi}_{L+G}=\hat{I}-(\hat{L}+\hat{G})_{R}^{-1}(\hat{L}+\hat{G})\label{eq:7.13}\end{equation}
}{\large \par}

{\large With condition (\ref{eq:5.6}), Eq.\ref{eq:7.12} leads to
an equation in which the permutation symmetry of solutions is explicitly
taken into account:}{\large \par}

{\large \begin{eqnarray}
 & |V>+\lambda_{1}\hat{S}(\hat{L}+\hat{G})_{R}^{-1}\cdot\nonumber \\
 & \left([\hat{I}+(\lambda_{2}\hat{M})_{R}^{-1}]^{-1}[(\lambda_{2}\hat{M})_{R}^{-1}+\hat{\Gamma}_{M}\hat{R}(\lambda_{2})]\right)\hat{N}|V>\nonumber \\
 & =\hat{S}\hat{\Pi}_{L+G}|V>+\nonumber \\
 & \hat{S}(\hat{L}+\hat{G})_{R}^{-1}\left(\hat{P}_{0}|V>+\lambda_{1}\hat{P}_{0}(\hat{I}+\lambda_{2}\hat{M})_{R}^{-1}\hat{N}|V>\right)\label{eq:7.14}\end{eqnarray}
}{\large \par}

{\large For a choice of (\ref{eq:8.8}), this equation in which the
arbitrary element - the projection $\hat{\Pi}_{L+G}|V>$- is the same
as in the theory with major coupling constant $\lambda_{1}=0$, see
Sec.5, can be closed, see the next section. It is noteworthy that
perhaps equations: (\ref{eq:7.4}), (\ref{eq:7.14}) are a new type
of equations which offer a new perspective on the calculation of n-pfs,
see Sec.10. It is encouraging that in both formulations unspecified
elements can be chosen in such a way that similar results can be obtained. }{\large \par}

{\large There is possibility of a different regularization of Eq.\ref{eq:4.5}:}{\large \par}

{\large \begin{eqnarray}
 & \left(\hat{L}+\lambda_{1}\hat{N}(\hat{I}+\lambda_{2}\hat{M})_{R}^{-1}+\hat{G}\right)|V>=\nonumber \\
 & \hat{P}_{0}|V>+\lambda_{1}\hat{P}_{0}\hat{N}(\hat{I}+\lambda_{2}\hat{M})_{R}^{-1}|V>\label{eq:7.15}\end{eqnarray}
 Hence and from (\ref{eq:7.8}) we get}{\large \par}

{\large \begin{eqnarray}
 & \left\{ \hat{L}++\hat{G}\right\} |V>+\nonumber \\
 & \lambda_{1}\hat{N}\left([\hat{I}+(\lambda_{2}\hat{M})_{R}^{-1}]^{-1}[(\lambda_{2}\hat{M})_{R}^{-1}+\hat{\Gamma}_{M}\hat{R}(\lambda_{2})]\right)|V>\nonumber \\
 & =\hat{P}_{0}|V>+\lambda_{1}\hat{P}_{0}\hat{N}(\hat{I}+\lambda_{2}\hat{M})_{R}^{-1}|V>\label{eq:7.16}\end{eqnarray}
 With choice (\ref{eq:8.8})}{\large \par}

{\large \begin{eqnarray}
 & |V>+(\hat{L}+\hat{G})_{R}^{-1}\cdot\nonumber \\
 & \left(\frac{\lambda_{1}}{\lambda_{2}}[\hat{I}+(\lambda_{2}\hat{N})_{R}^{-1}]^{-1}+\lambda_{1}\hat{N}\left([\hat{I}+(\lambda_{2}\hat{N})_{R}^{-1}]^{-1}\hat{\Gamma}_{N}\hat{R}(\lambda_{2})]\right)\right)\cdot\nonumber \\
 & \cdot|V>=\hat{\Pi}_{L+G}|V>+\nonumber \\
 & (\hat{L}+\hat{G})_{R}^{-1}\left(\hat{P}_{0}|V>+\lambda_{1}\hat{P}_{0}\hat{N}(\hat{I}+\lambda_{2}\hat{N})_{R}^{-1}|V>\right)\label{eq:7.18}\end{eqnarray}
 With assumption (\ref{eq:5.6}) the above equation leads to its symmetrized
version}{\large \par}

{\large \begin{eqnarray}
 & |V>+\hat{S}(\hat{L}+\hat{G})_{R}^{-1}\cdot\nonumber \\
 & \left(\frac{\lambda_{1}}{\lambda_{2}}[\hat{I}+(\lambda_{2}\hat{N})_{R}^{-1}]^{-1}+\lambda_{1}\hat{N}\left([\hat{I}+(\lambda_{2}\hat{N})_{R}^{-1}]^{-1}\hat{\Gamma}_{N}\hat{R}(\lambda_{2})]\right)\right)\cdot\nonumber \\
 & \cdot|V>=\hat{S}\hat{\Pi}_{L+G}|V>+\nonumber \\
 & \hat{S}(\hat{L}+\hat{G})_{R}^{-1}\left(\hat{P}_{0}|V>+\lambda_{1}\hat{P}_{0}\hat{N}(\hat{I}+\lambda_{2}\hat{N})_{R}^{-1}|V>\right)\nonumber \\
\label{eq:7.19}\end{eqnarray}
}{\large \par}

{\large This equation as other previous equations, for the choice
(\ref{eq:8.8}), leads to closed equations for n-pfs. A positive fact
is that, as in perturbation theory, the undetermined vector $\hat{S}\hat{\Pi}_{L+G}|V>$can
be identified with the original linear theory ($\hat{N}=0$) and that
for derivation of the last equation we did not project, besides their
symmetrization, the previous equations. This last fact makes that
weakening of the original Eqs (\ref{eq:2.1}) still can be interpreted
as the appropriate averaging. }{\large \par}

{\large The arbitrary term, $\hat{\Gamma}_{N}\hat{R}(\lambda_{2})$,
can be chosen in such a way to remove the {}``secular'' terms which
arise when the $\lambda_{2}\rightarrow0$.}{\large \par}

{\large Finally, we can connect two kind of regularizations and consider
the following regularization:}{\large \par}

{\large \begin{equation}
\hat{N}\rightarrow\frac{1}{2}\left(\hat{N}(\hat{I}+\lambda_{2}\hat{N})_{R}^{-1}+(\hat{I}+\lambda_{2}\hat{N})_{R}^{-1}\hat{N}\right)\label{eq:7.20}\end{equation}
to mimic in some degree a commutation of $\hat{N}$ with operator
$(\hat{I}+\lambda_{2}\hat{N})^{-1}$if the last operator would exist. }{\large \par}

{\large For such regularization and choice (\ref{eq:7.10}) chosen
only for simplicity, we get the following equation: }{\large \par}

{\large \begin{eqnarray}
 & \left\{ \hat{I}+\frac{\lambda_{1}}{2\lambda_{2}}\hat{S}(\hat{L}+\hat{G})_{R}^{-1}\left([\hat{I}+(\lambda_{2}\hat{N})_{R}^{-1}]^{-1}[\hat{Q}_{N}+\hat{I}]\right)\right\} \cdot\nonumber \\
 & \cdot|V>=\hat{S}\hat{\Pi}_{L+G}|V>+\hat{S}(\hat{L}+\hat{G})_{R}^{-1}\cdot\nonumber \\
 & \left(\hat{P}_{0}|V>+\frac{\lambda_{1}}{2}\hat{P}_{0}\left(\hat{N}(\hat{I}+\lambda_{2}\hat{N})_{R}^{-1}+(\hat{I}+\lambda_{2}\hat{N})_{R}^{-1}\hat{N}\right)|V>\right)\nonumber \\
\label{eq:7.21}\end{eqnarray}
}{\large \par}

{\large We see that at the diagonal terms the rate of coupling constants
$\lambda_{1,2}$appears and particularly, the 1-pf V only has such
a dependence. This means that, for monomial theories at least and
used assumptions, the 1-pfs V, in the $lim\lambda_{2}\rightarrow0$,
stop to depend on the major coupling constant $\lambda_{1}$. If we
agree that exact closure of equations for n-pfs V is equivalent to
a summation of the infinite terms of the canonical perturbation theory,
we can compare this result with disappearing of the effective coupling
constant of the perturbation theory. This rather negative result can
be treated as a hint that $\lambda_{2}\neq0$, and that non polynomial
theories should be taken into consideration, or, that $\lambda_{1}$depends
on $\lambda_{2}$in such a way that $\frac{\lambda_{1}}{\lambda_{2}}\rightarrow const$
when $\lambda_{2}\rightarrow0$ (some kind of renormalization of the
original theory). We can imagine the following situation in which
the developed type of theories can be used: For a certain class of
initial and/or boundary conditions ({}``laminar'' conditions), phenomena
are well described by the polynomial theory (\ref{eq:2.1}) with $\lambda_{2}=0$.
For others ({}``turbulent'' conditions, we need to use averages
or smearing and then non polynomial terms ($\lambda_{2}\neq0$) have
to be taken into account because, as we see in averaged version of
Eqs (\ref{eq:2.1}), Eqs (\ref{eq:7.1}) and (\ref{eq:7.2}), $\lambda_{2}$enters
these equations in the inverse powers. In other words, the non polynomial
terms become apparent only when considered situations (boundary and
initial conditions) are becoming increasingly complex and averaged
solutions and correlation functions are needed. }{\large \par}

{\large Let us show that at least for a choice (\ref{eq:7.10}) the
right invertible operator $\hat{M}$ dose not commute with its general
resolvent $\hat{R}=[\hat{I}+(\lambda\hat{M})_{R}^{-1}]^{-1}(\lambda\hat{M})_{R}^{-1}$:
We get}{\large \par}

{\large \begin{eqnarray}
 & \left[\hat{M},\hat{R}\right]=\hat{M}\frac{(\lambda\hat{M})_{R}^{-1}}{\hat{I}+(\lambda\hat{M})_{R}^{-1}}-\frac{(\lambda\hat{M})_{R}^{-1}}{\hat{I}+(\lambda\hat{M})_{R}^{-1}}\hat{M}=\nonumber \\
 & \frac{\lambda^{-1}\hat{I}}{\hat{I}+(\lambda\hat{M})_{R}^{-1}}-\frac{\lambda^{-1}\hat{Q}_{M}}{\hat{I}+(\lambda\hat{M})_{R}^{-1}}=\nonumber \\
 & \frac{\lambda^{-1}\hat{\Gamma}_{M}}{\hat{I}+(\lambda\hat{M})_{R}^{-1}}=\lambda^{-1}\hat{\Gamma}_{M}\label{eq:7.22}\end{eqnarray}
see (\ref{eq:7.6}) and (\ref{eq:7.9}), for definitions of appropriate
projectors. }{\large \par}

\section{{\large Undetermined terms of the general solution. }}

{\large If we confine ourselves to the solutions satisfying the condition
(\ref{eq:5.6}), then the projected by the projector $\hat{S}$, Eq.\ref{eq:7.4},
can be written as follows:}{\large \par}

{\large \begin{eqnarray}
 & \left\{ \hat{I}+\hat{S}[\lambda_{2}\hat{M}(\hat{L}+\hat{G})]_{R}^{-1}[\hat{L}+\hat{G}+\lambda_{1}\hat{N}]\right\} |V>=\hat{S}\hat{P}_{M}|V>+\nonumber \\
 & \hat{S}[\lambda_{2}\hat{M}(\hat{L}+\hat{G})]_{R}^{-1}\left(\hat{P}_{0}|V>+\lambda_{1}\hat{P}_{0}(\hat{I}+\lambda_{2}\hat{M})_{R}^{-1}\hat{N}|V>\right)\nonumber \\
\label{eq:8.1}\end{eqnarray}
}{\large \par}

{\large Introducing the operator}{\large \par}

{\large \begin{equation}
\hat{A}=\hat{I}+\hat{S}[\lambda_{2}\hat{M}(\hat{L}+\hat{G})]_{R}^{-1}[\hat{L}+\hat{G}]\label{eq:8.2}\end{equation}
we rewrite Eq.\ref{eq:8.1} as follows:}{\large \par}

{\large \begin{eqnarray}
 & \{\hat{A}+\lambda_{1}\hat{S}[\lambda_{2}\hat{M}(\hat{L}+\hat{G})]_{R}^{-1}\hat{N}\}|V>=\hat{S}\hat{P}_{M}|V>+\nonumber \\
 & \hat{S}[\lambda_{2}\hat{M}(\hat{L}+\hat{G})]_{R}^{-1}\left(\hat{P}_{0}|V>+\lambda_{1}\hat{P}_{0}(\hat{I}+\lambda_{2}\hat{M})_{R}^{-1}\hat{N}|V>\right)\nonumber \\
\label{eq:8.3}\end{eqnarray}
This equation may serve as a starting point for a perturbation theory
of the modified Eq\ref{eq:7.1}with the major coupling constant $\lambda_{1}$as
an expansion parameter and the operator $\hat{S}[\lambda_{2}\hat{M}(\hat{L}+\hat{G})]_{R}^{-1}\hat{N}$
as the perturbation operator. With assumption about non singularity
of the operator $\hat{A}$, we can rewrite Eq.\ref{eq:8.3-1} as }{\large \par}

{\large \begin{eqnarray}
 & \left\{ \hat{I}+\lambda_{1}\hat{A}^{-1}\hat{S}[\lambda_{2}\hat{M}(\hat{L}+\hat{G})]_{R}^{-1}\hat{N}\right\} |V>=\hat{A}^{-1}\hat{S}\hat{P}_{M}|V>+\nonumber \\
 & \hat{A}^{-1}\hat{S}[\lambda_{2}\hat{M}(\hat{L}+\hat{G})]_{R}^{-1}\left(\hat{P}_{0}|V>+\lambda_{1}\hat{P}_{0}(\hat{I}+\lambda_{2}\hat{M})_{R}^{-1}\hat{N}|V>\right)\nonumber \\
\label{eq:8.4}\end{eqnarray}
 We will assume that, for $\lambda_{1}=0$, the theory is continuous
and hence, the unknown term}{\large \par}

{\large \begin{eqnarray}
 & \hat{A}^{-1}\hat{S}\hat{P}_{M}|V>^{(0)}+\hat{A}^{-1}\hat{S}[\lambda_{2}\hat{M}(\hat{L}+\hat{G})]_{R}^{-1}\hat{P}_{0}|V>^{(0)} & =\nonumber \\
 & |V;\lambda_{1}=0>\equiv|V>^{(0)}\label{eq:8.5}\end{eqnarray}
where the generating vector $|V>^{(0)}$generates correlation functions
of the originally linear theory characterized by the zero coupling
constant $\lambda_{1}$. In a general case , for $\lambda_{1}\ncong0$,
the undetermined element of Eq.\ref{eq:8.4} is chosen in a such way
that the IBC are satisfied. Additionally, if the expansion (\ref{eq:5.9})
is used and we get}{\large \par}

{\large \begin{eqnarray}
 & \hat{A}^{-1}\hat{S}\hat{P}_{M}|V;\lambda_{1},\lambda_{2}>+\hat{A}^{-1}\hat{S}[\lambda_{2}\hat{M}(\hat{L}+\hat{G})]_{R}^{-1}\hat{P}_{0}|V> & =\nonumber \\
 & |V>^{(0)}+\sum_{j=1}^{\propto}\lambda_{1}^{j}(\hat{A}^{-1}\hat{S}\hat{P}_{M}|V;\lambda_{2}>)^{(j)}\label{eq:8.6}\end{eqnarray}
then we try to choose the undetermined projections $\hat{S}\hat{P}_{M}|V;\lambda_{2}>^{(j)}$
in such a way that no divergent terms do not appear in the series
(\ref{eq:5.9}), when $\lambda_{2}\rightarrow0$. We have assumed
here that the original theory ($\lambda_{2}=0$) is well defined. }{\large \par}

{\large The simplest case is then again when}{\large \par}

{\large \begin{eqnarray}
 & \hat{A}^{-1}\hat{S}\hat{P}_{M}|V>+\hat{A}^{-1}\hat{S}[\lambda_{2}\hat{M}(\hat{L}+\hat{G})]_{R}^{-1}\hat{P}_{0}|V>\nonumber \\
 & =|V>^{(0)}\Leftrightarrow\nonumber \\
 & \hat{S}\hat{P}_{M}|V>+\hat{S}[\lambda_{2}\hat{M}(\hat{L}+\hat{G})]_{R}^{-1}\hat{P}_{0}|V>=\hat{A}(\lambda_{2})|V>^{(0)}\nonumber \\
\label{eq:8.7}\end{eqnarray}
This assumption is in agreement with (\ref{eq:8.5}) and also corresponds
to the case when the IBC are satisfied by the zero-order approximation
and no divergences appear in the expansion with respect to the major
coupling constant $\lambda_{1}$. The assumption that arbitrary term
of Eq.\ref{eq:8.4}, $\hat{A}^{-1}\hat{S}\hat{P}_{M}|V>$, is identified
with the original linear theory, see Eq.\ref{eq:2.1}, indicates that
the further consequences of this theory are related to the nonlinear
effects (perturbation of the original, linear theory, (Eq.\ref{eq:2.1}))
occurring in the system. Quite natural assumption which is a started
point of almost every perturbation theory. Again we must recall that
assumption (\ref{eq:8.7}) does not exclude that the part of arbitrary
element $\hat{P}_{M}|V>\in\hat{P}_{M}F$ }{\large \par}

{\large \begin{equation}
(\hat{I}-\hat{S})\hat{P}_{M}|V>=f(\lambda_{1},\lambda_{2})\label{eq:8.7'}\end{equation}
responsible for the elimination of asymmetric parts in Eq.\ref{eq:7.4}
can be a complicated vector function allowing to satisfy the condition
(\ref{eq:5.6}). }{\large \par}

{\large If at the choice (\ref{eq:8.7}) some bad, {}``secular''
terms appear in the power series expansion with respect to the major
coupling constant $\lambda_{1}$, see \cite{Nayfeh (1972)} then to
remove such terms we can use the formula (\ref{eq:8.6}). It appears
an open question: Is it possible using this type of equations to gain
a new perspective on the renormalization in of quantum and statistical
field theories?}{\large \par}

{\large Next we assume}{\large \par}

{\large \begin{equation}
\hat{M}=\hat{N}\label{eq:8.8}\end{equation}
because such a choice of the modified operator often leads to closed
or partly closed equations for n-point correlation functions. It is
natural that a candidate for modified theory should be looked among
closed theories of the type (\ref{eq:8.8}). It is worth noting that
a more general choice of the modifying operator }{\large \par}

{\large \begin{equation}
\hat{M}=\hat{\Lambda}\hat{N}\label{eq:8.9}\end{equation}
 with a diagonal operator $\hat{\Lambda}$ can also be used to close
the considered equations. We assume that operator $\hat{\Lambda}$is
a non singular operator not to increase freedom of the theory contained
in the projection $\hat{\Gamma}_{M}\hat{R}(\lambda_{2})$, see (\ref{eq:7.8}).
Nevertheless, the choice of (\ref{eq:8.8}) exhibits a kind of self-similarities
of fractal theories. }{\large \par}

\section{{\large Functional calculus and considered equations}}

{\large The }\textit{\large functional calculus}{\large{} is define
sometimes as a branch of mathematics about inserting operators into
functions to get in result meaningful new operators, see, e.g., \cite{Haas (2007)},
\cite{Internet (2010)}. Such operator-valued functions often appear
in solutions to many linear equations of physics and engineering problems
and in our opinion are an important generalization of the concept
of function. }{\large \par}

{\large From the functional calculus point of view we can say that
our paper needs a definition of the function $(\hat{I}+\lambda_{2}\hat{M})^{-1}$in
the case of upper triangular operators $\hat{M}$ like (\ref{eq:8.8}).
The natural and often used definition of this function as the power
series}{\large \par}

{\large \begin{equation}
(\hat{I}+\lambda_{2}\hat{M})^{-1}:=\hat{I}-\lambda_{2}\hat{M}+(\lambda_{2}\hat{M})^{2}-...\label{eq:9.1}\end{equation}
for unbounded operators, in every its term can be incorrect and, moreover,
in the case of upper triangular operators $\hat{M}$ very inconvenient
because it introduces to every correlation function of basic Eq\ref{eq:7.1}
an infinite number of modification or regularization terms. Therefore,
we used a different definition}{\large \par}

{\large \begin{equation}
(\hat{I}+\lambda_{2}\hat{M})^{-1}:=(\hat{I}+\lambda_{2}\hat{M})_{R}^{-1}\label{eq:9.2}\end{equation}
 where subscript {}``R'' means a right inverse operator to the operator
$\hat{I}+\lambda_{2}\hat{M}$. So, instead of a senseless expression
(\ref{eq:9.1}) we use well defined and explicitly constructed expression
(\ref{eq:9.2}), see (\ref{eq:7.8}), (\ref{eq:8.8}) and (\ref{eq:10.3})-(\ref{eq:10.3''}).
The operator (\ref{eq:9.2}) is called the }\textit{\large generalized
resolvent}{\large{} of the operator $\hat{M}$. Interesting thing is
that regularization or modification of equations (\ref{eq:4.5}),
given by the generalized resolvent of the operator $\hat{M}=\hat{N}$,
leads to closed equations for the correlation functions. Moreover,
many operator-valued functions in the free Fock space can be defined
by means of the type of Dunford-Cauchy integral with generalized resolvent
(\ref{eq:9.2}):}{\large \par}

{\large \begin{equation}
f_{\lambda_{2}}(\hat{M})\equiv\intop_{\Gamma}f_{\lambda_{2}}(\lambda)(\hat{I}+\lambda\hat{M})_{R}^{-1}d\lambda\label{eq:9.3}\end{equation}
}{\large \par}

{\large It is not unlikely that this paper is the first step to understanding
these operator valued functions in the context of equations for the
correlation functions. We can say that we analyzed such a problem:
to which conclusions leads a regularization or modification of Eq.\ref{eq:4.5}
which consists in replacing an upper triangular operator $\hat{N}$
by the diagonal + lower triangular operator, e.g.,}{\large \par}

{\large \begin{equation}
\hat{N}\rightarrow f_{\lambda_{2}}(\hat{N})\hat{N}=\left(\intop_{\Gamma}f_{\lambda_{2}}(\lambda)[\hat{I}+(\lambda\hat{N})_{R}^{-1}]^{-1}(\lambda N)_{R}^{-1}d\lambda\right)\hat{N}\label{eq:9.4}\end{equation}
or }{\large \par}

{\large \begin{equation}
\hat{N}\rightarrow\frac{1}{2}\left(f_{\lambda_{2}}(\hat{N})\hat{N}+\hat{N}f_{\lambda_{2}}(\hat{N})\right)\label{eq:9.5}\end{equation}
see (\ref{eq:7.8}), (\ref{eq:7.10}) and (\ref{eq:8.8}). In the
last formula like in (\ref{eq:7.20}) we took into account that different
right inverse operators do not commute with each other: Denoting two
different right inverses by $\hat{N}_{R}^{-1}$and $\hat{N}_{R}^{-1'}$if
we assume their commutativity we come to the absurd:}{\large \par}

{\large \begin{equation}
\hat{N}([\hat{N}_{R}^{-1},\hat{N}_{R}^{-1'}])|\Phi>=\hat{N}_{R}^{-1'}|\Phi>-\hat{N}_{R}^{-1}|\Phi>=0\end{equation}
 for any vector $|\Phi>$. Also the operator valued functions constructed
by different right inverses do not commute with each other and the
original operator $\hat{N}$. For example, we expect that, at least
formally, $\hat{N}$commute with operator valued functions like $(\hat{I}+\lambda\hat{N})^{-1}$or
$exp(-\hat{N})$ and so on. Nevertheless, we try to mimic commutativity
of their formal predecessors (formal inverse operators, see (\ref{eq:9.1})).
We can support this idea in the following way: in the case in which
the problem in consideration is ill-posed and we use averages, then
the original Eq.\ref{eq:2.1} looses its meaning and a more free connection
between averages and Eq.\ref{eq:2.1} is recommended. In this way
we substitute traditional theories with very complicated perturbation
series and related to the open equations for correlation functions
by a new, more flexible theories with closed equations like (\ref{eq:7.21-1}). }{\large \par}

{\large In the case of regularization of the original theory with
nonlinearity described by the operator $\hat{N}$ we have a demand}{\large \par}

{\large \begin{equation}
lim_{\lambda_{2}\rightarrow0}f_{\lambda_{2}}(\hat{N})=\hat{I}\label{eq:9.6}\end{equation}
 or its weaker form, e.g., }{\large \par}

{\large \begin{equation}
lim_{\lambda_{2}\rightarrow0}f_{\lambda_{2}}(\hat{N})=\hat{I}-\hat{P}_{0}\label{eq:9.7}\end{equation}
It is not clear whether the condition (\ref{eq:9.6}) or (\ref{eq:9.7}),
although weaker than the condition (\ref{eq:7.10'}) and (\ref{eq:7.10''}),
can be fulfilled for the operator valued functions occurring in formulas
(\ref{eq:9.4}) and (\ref{eq:9.5}). }{\large \par}

\section{{\large An example of the theory with a local nonlinear term, divergences
and final remarks}}

{\large Let us consider the equation:}{\large \par}

{\large \begin{equation}
\int L(\tilde{x},\tilde{y})\varphi(\tilde{y})d\tilde{y}+\lambda\varphi^{3}(\tilde{x})=0\label{eq:10.1}\end{equation}
 which is called the $\varphi^{(3)}$- or Hurst model. In this model
the function $G=0$ in Eq.\ref{eq:2.1}. Usually, the kernel $L$
is a sum of the Dirac's deltas and its derivatives or their discrete
version. For this model, the operator }{\large \par}

{\large \begin{equation}
\hat{L}=\int\hat{\eta}*(\tilde{x})L(\tilde{x},\tilde{y})\hat{\eta}(\tilde{y})d\tilde{x}d\tilde{y}+\hat{P}_{0}\label{eq:10.2}\end{equation}
 the operator }{\large \par}

{\large \begin{equation}
\hat{N}=\int\hat{\eta}*(\tilde{z})\hat{\eta}(\tilde{z})^{2}d\tilde{z}+\hat{P}_{0}\hat{N}\label{eq:10.3}\end{equation}
where, to guarantee right invertibility of the operator $\hat{N}$,
we have chosen}{\large \par}

{\large \begin{equation}
\hat{P}_{0}\hat{N}=\hat{P}_{0}\hat{N}\hat{P}_{1}=\hat{P}_{0}\int d\tilde{z}\hat{\eta}(\tilde{z})d\tilde{z}\hat{P}_{1}\neq0\label{eq:10.3'}\end{equation}
In this case a right inverse to the operator $\hat{N}$ can be chosen
as }{\large \par}

{\large \begin{equation}
\hat{N}_{R}^{-1}=\int\{\hat{\eta}*(\tilde{y}))^{2}\hat{\eta}(\tilde{y})d\tilde{y}+\hat{N}_{R}^{-1}\hat{P}_{0}\label{eq:10.3''}\end{equation}
with }{\large \par}

{\large \begin{equation}
\hat{N}_{R}^{-1}\hat{P}_{0}=\hat{P}_{1}\hat{N}_{R}^{-1}\hat{P}_{0}=\hat{P}_{1}\hat{\eta}^{\star}(\tilde{y})\hat{P}_{0}\label{eq:10.3'''}\end{equation}
}{\large \par}

{\large We will use the regularized theory leading to closed equations,
Secs 7-8. For a sake of simplicity, we use Eq.\ref{eq:7.14} in the
case of (\ref{eq:8.8}) and the operator $\hat{G}=0$. We also choose
(\ref{eq:7.10}), for the arbitrary element of the generalized resolvent
(\ref{eq:7.8}). In result, we get equation}{\large \par}

{\large \begin{eqnarray}
 & \left\{ \hat{I}+\lambda_{1}\hat{S}(\hat{L})_{R}^{-1}\left([\hat{I}+(\lambda_{2}\hat{N})_{R}^{-1}]^{-1}(\lambda_{2}\hat{N})_{R}^{-1}\right)\hat{N}\right\} |V>=\nonumber \\
 & \hat{S}\hat{\Pi}_{L}|V>+\hat{S}(\hat{L})_{R}^{-1}\left(\hat{P}_{0}|V>+\lambda_{1}\hat{P}_{0}(\hat{I}+\lambda_{2}\hat{N})_{R}^{-1}\hat{N}|V>\right)\nonumber \\
\label{eq:10.4}\end{eqnarray}
and further}{\large \par}

{\large \begin{eqnarray}
 & \left\{ \hat{I}+\frac{\lambda_{1}}{\lambda_{2}}\hat{S}(\hat{L})_{R}^{-1}\left([\hat{I}+(\lambda_{2}\hat{N})_{R}^{-1}]^{-1}\hat{Q}_{N}\right)\right\} |V>=\hat{S}\hat{\Pi}_{L}|V>\nonumber \\
 & +\hat{S}(\hat{L})_{R}^{-1}\left(\hat{P}_{0}|V>+\lambda_{1}\hat{P}_{0}(\hat{I}+\lambda_{2}\hat{N})_{R}^{-1}\hat{N}|V>\right)\nonumber \\
\label{eq:10.5}\end{eqnarray}
where projector}{\large \par}

{\large \begin{equation}
\hat{Q}_{N}=\hat{N}_{R}^{-1}\hat{N}\label{eq:10.6}\end{equation}
 If projection $\hat{S}\hat{\Pi}_{L}|V>$can be calculated from Eq.\ref{eq:10.5}
in which $\lambda_{1}=0$ we would get}{\large \par}

{\large \begin{eqnarray}
\hat{S}\hat{\Pi}_{L}|V & > & =\{\hat{I}-\hat{S}\hat{L}_{R}^{-1}\hat{P}_{0}\}|V>^{(0)}=\{\hat{I}-\hat{P}_{0}\}|V>^{(0)}\nonumber \\
\label{eq:10.7}\end{eqnarray}
}{\large \par}

{\large From Eq.\ref{eq:10.5} we get then}{\large \par}

{\large \begin{eqnarray}
 & \hat{P}_{2}\{\hat{I}+\frac{\lambda_{1}}{\lambda_{2}}\hat{S}\hat{L}_{R}^{-1}\left(\hat{I}-\lambda_{2}^{-1}\hat{N}_{R}^{-1}\right)\hat{Q}_{N}\}|V>=\nonumber \\
 & \hat{P}_{2}\hat{S}\hat{\Pi}_{L}|V>=\hat{P}_{2}|V>^{(0)}\label{eq:10.8}\end{eqnarray}
 and }{\large \par}

{\large \begin{eqnarray}
\hat{P}_{1}\{\hat{I}+\frac{\lambda_{1}}{\lambda_{2}}\hat{S}\hat{L}_{R}^{-1}\hat{Q}_{N}\}|V & > & =\hat{P}_{1}\hat{S}\hat{\Pi}_{L}|V>=\hat{P}_{1}|V>^{(0)}\nonumber \\
\label{eq:10.9}\end{eqnarray}
 where we took into account that the operator $\hat{N}_{R}^{-1}$is
lower triangular and }{\large \par}

{\large \begin{equation}
\hat{P}_{0}\hat{Q}_{N}=0\label{eq:10.9'}\end{equation}
}{\large \par}

{\large These are closed equations for the lowest - 1 and 2-pfs. We
can't get an equation for the 0-pf, $V_{0}$, see (\ref{eq:4.1}).
With Eq.\ref{eq:10.9} we can express the 1-point solution as:}{\large \par}

{\large \begin{equation}
\hat{P}_{1}|V>=\hat{P}_{1}\{\hat{I}+\frac{\lambda_{1}}{\lambda_{2}}\hat{S}\hat{L}_{R}^{-1}\hat{Q}_{N}\}^{-1}|V>^{(0)}\label{eq:10.9''}\end{equation}
}{\large \par}

{\large To describe explicitly these equations we have to take into
account that }{\large \par}

{\large \begin{eqnarray}
 & \hat{Q}_{N}=\hat{N}_{R}^{-1}\hat{N}=\nonumber \\
 & \int\hat{\eta}^{\star}(\tilde{x})^{2}\hat{\eta}(\tilde{x})^{2}d\tilde{x}+\hat{P}_{1}\int\hat{\eta}^{\star}(\tilde{z})\hat{P}_{0}\hat{\eta}(\tilde{y})d\tilde{y}\hat{P}_{1}\label{eq:10.10}\end{eqnarray}
 and }{\large \par}

{\large \begin{eqnarray}
 & \hat{L}_{R}^{-1}\hat{Q}_{N}=\int\hat{\eta}^{\star}(\tilde{x})L_{R}^{-1}(\tilde{x},\tilde{y})\hat{\eta}^{\star}(\tilde{y})\hat{\eta}(\tilde{y})^{2}d\tilde{x}d\tilde{y}+\nonumber \\
 & \hat{P}_{1}\hat{L}_{R}^{-1}\hat{P}_{1}\int\hat{\eta}^{\star}(\tilde{z})\hat{P}_{0}\hat{\eta}(\tilde{y})d\tilde{y}\hat{P}_{1}\nonumber \\
\label{eq:10.11}\end{eqnarray}
 with arbitrary value for the variable $\tilde{z}$. From (\ref{eq:10.9})
we get the following equation for the 1-pf:}{\large \par}

{\large \begin{equation}
V(\tilde{x})+\frac{\lambda_{1}}{\lambda_{2}}\int L_{R}^{-1}(\tilde{x},\tilde{y})d\tilde{y}\, V(\tilde{z})=V^{(0)}(\tilde{x})\label{eq:10.12}\end{equation}
 with arbitrary fixed value for the variable $\tilde{z}$ . Denoting
the integral }{\large \par}

{\large \begin{equation}
\int L_{R}^{-1}(\tilde{x},\tilde{y})d\tilde{y}=a(\tilde{x})\label{eq:10.13}\end{equation}
 we rewrite Eq.\ref{eq:10.12} as}{\large \par}

{\large \begin{equation}
V(\tilde{x})+\frac{\lambda_{1}}{\lambda_{2}}a(\tilde{x})V(\tilde{z})=V^{(0)}(\tilde{x})\label{eq:10.14}\end{equation}
 Its solution is }{\large \par}

{\large \begin{equation}
V(\tilde{x})\equiv V(\tilde{x};\tilde{z})=V^{(0)}(\tilde{x})-\frac{\lambda_{1}}{\lambda_{2}}a(\tilde{x})\frac{V^{(0)}(\tilde{z})}{1+\frac{\lambda_{1}}{\lambda_{2}}a(\tilde{z})}\label{eq:10.15}\end{equation}
 The result is rather frustrating because, for $\lambda_{2}\rightarrow0$,
dependence of the 1-pf V on $\lambda_{1}$ disappears completely.
Similar results we get for the 2-pf with trivial condition (\ref{eq:3.3}),
for $\lambda_{2}\rightarrow0$. This could mean that perhaps we should
keep $\lambda_{2}\neq0$ , or, we should try to use other regularization
also proposed in Sec.7. }{\large \par}


\begin{thebibliography}{21}
{\large \bibitem[1]{Noak (2010)} Noack, B.R., M.Schlegel, M.Morzy\'{n}ski
aasnd G.Tadmor. 2010. }\textit{\large System reduction strategy for
Galerkin models of fluid flows. }{\large Int. J. Numer. Meth. Fluids
}\textbf{\large 63}{\large , 231-248}{\large \par}

{\large \bibitem[2]{Cordier (2009)} Cordier, L., B.Abou El Majd and
J.Favier. 2009. }\textit{\large Calibration of POD Reduced-Order Model
using Tikhonov regularization.}{\large{} Int. J. Numer. Meth. Fluids
}\textbf{\large 00, }{\large 1-00. }{\large \par}

{\large \bibitem[3]{Montlaur (2009)} de Montlaur, A. 2009. }\textit{\large High-Order
Discontinuous Galerkin Methods for Incompressible Flows. }{\large Internet}{\large \par}

{\large \bibitem[4]{Cookburn (2003)} Cookburn, B. 2003. }\textit{\large Discontinuous
Galerkin Methods. }{\large School of Mathematics, Univ. of Minnesota
1-25. Internet }{\large \par}

{\large \bibitem[5]{Gunzburger (????)} Gunzburger, M. ????. }\textit{\large Reduced-Order
Modeling. }{\large Florida State University. Internet }{\large \par}

{\large \bibitem[6]{Hanckow (2007)} Hanckowiak, J. 2007. }\textit{\large Reylnolds'
Dream? }{\large Nonlinear Dyn. }\textbf{\large 50, }{\large 191-211}{\large \par}

{\large \bibitem[7]{Hanckow (2008)} Hanckowiak, J. 2008. }\textit{\large Some
Insight into Many Constituent Dynamics. }{\large arXiv:0807.1489 (July
2008). Internet}{\large \par}

{\large \bibitem[8]{Hanckow (2009)} Hanckowiak, J. 2009. }\textit{\large Physics
Equations with Closure Principle. }{\large Preprint of Zielona Gora
University}{\large \par}

{\large \bibitem[9]{Hanckow (2010)} Hanckowiak, J. 2010. }\textit{\large Models
of the {}``Universe'' and a Closure Principle. }{\large http://arxiv.org/submit/127988}{\large \par}

{\large \bibitem[10]{Barnaby (2008)} Barnaby, N. and N.Kamran. 2008.
}\textit{\large Dynamics with Infinitely Many Derivatives. The Initial
Value Problem. }{\large arXiv.:0709.3968v3}{\large \par}

{\large \bibitem[11]{Kato (1966)} Kato, T. 1966. }\textit{\large Perturbation
theory for linear operators. }{\large Springer-Verlag Berlin- Heidelberg-New
York}{\large \par}

{\large \bibitem[12]{Badino (2005)}Badino, M. 2005. }\textit{\large The
Foundational Role of Ergodic Theory. }{\large Max-Planck Institute
for the History of Science, Preprint 292. Internet}{\large \par}

{\large \bibitem[13]{Svierish (1996)}Sviershchevskii, S. R. 1996.
}\textit{\large Invariant Linear Spaces ans Exact Solutions of Nonlinear
Evolution Equations. }{\large Nonlinear Math. Phys. v. 3, N 1-2, pp
164-169 }{\large \par}

{\large \bibitem[14]{Hydon (2005)}Hydon, P.E. 2005. }\textit{\large An
introduction to symmetry methods in the solution of differential equations
that occur in chemistry and chemical biology. }{\large Internet}{\large \par}

{\large \bibitem[15]{Christensen (1999)}Christensen, O. 1999. }\textit{\large Operators
with Closed Range, Pseudo-Inverse, and Perturbation of Frames for
a Subspace. }{\large Canad. Math. Bull. 42 (1), pp. 37-45 (Internet)}{\large \par}

{\large \bibitem[16]{Salam et al. (1971)}Salam, A. and J. Strathdee.
1971. Triest preprint IC/71/14 (Internet)}{\large \par}

{\large \bibitem[17]{Kosyakov (2000)}Kosyakov, B.P. 2000. }\textit{\large Physical
sense of renormalizability. }{\large arXiv:hep-th/0011235v1}{\large \par}

{\large \bibitem[18]{Nayfeh (1972)}Nayfeh, A.H. 1972. }\textit{\large Perturbation
methods. }{\large A Wiley-Interscience Publication, John Wiley \&
Sons, New York - London - Sydney - Toronto}{\large \par}

{\large \bibitem[19]{Haas (2007)}Haas, M. 2007. }\textit{\large Functional
calculus for groups and applications to evolution equations. }{\large J.evol.equ.
7, 529-554}{\large \par}

{\large \bibitem[20]{Internet (2010)}2010. }\textit{\large Borel
functional calculus. wikipedia }{\large \par}

\bibitem[21]{Ivan (1974)}{\large S. S. Ivanov, D. Ya. Petrina and
A. L. Rebenko, 1974, }\textit{\large On equations for the coefficient
functions of the S matrix in quantum field theory. }{\large Theor.
Math. Phys., Vol. 19, No. 1, 332-339, }
\end{thebibliography}
\end{document}